\documentclass[AMA,STIX2COL]{MRM}
\articletype{PREPRINT - SUBMITTED TO MAGNETIC RESONANCE IN MEDICINE}%

\received{\today}
\revised{XXX}
\accepted{XXX}
\usepackage{color}
\usepackage{bbm}

\newcommand{\tstar}[1]{{T$_2^*$}}
\usepackage{pdfpages}

\begin{document}
%TC:ignore
\title{Motion-Robust \tstar{} Quantification from Gradient Echo MRI with Physics-Informed Deep Learning}

\author[1,2]{Hannah Eichhorn}{\orcid{0000-0001-6980-9703}}

\author[1,2]{Veronika Spieker}{\orcid{0000-0001-7720-7569}}

\author[2]{Kerstin Hammernik}{\orcid{0000-0002-2734-1409}}

\author[3,4]{Elisa Saks}{\orcid{0009-0002-2766-8054}}

\author[2]{Lina Felsner}{\orcid{0000-0001-7695-2612}} 

\author[5]{Kilian Weiss}{\orcid{0000-0003-4295-4585}}

\author[3,4,6]{Christine Preibisch}{\orcid{0000-0003-4067-1928}}

\author[1,2,7]{Julia A. Schnabel}{\orcid{0000-0001-6107-3009}}

\authormark{HANNAH EICHHORN \textsc{et al}}

\address[1]{\orgdiv{Institute of Machine Learning in Biomedical Imaging}, \orgname{Helmholtz Munich}, \orgaddress{\country{Germany}}}

\address[2]{\orgdiv{School of Computation, Information~\& Technology}, \orgname{Technical University of Munich}, \orgaddress{\country{Germany}}}

\address[3]{\orgdiv{School of Medicine \& Health, Institute for Diagnostic and Interventional Neuroradiology}, \orgname{Technical University of Munich}, \orgaddress{\country{Germany}}}

\address[4]{\orgdiv{School of Medicine \& Health, TUM-Neuroimaging Center}, \orgname{Technical University of Munich}, \orgaddress{\country{Germany}}}

\address[5]{\orgname{Philips GmbH Market DACH}, \orgaddress{\country{Germany}}}

\address[6]{\orgdiv{School of Medicine \& Health, Clinic of Neurology}, \orgname{Technical University of Munich}, \orgaddress{\country{Germany}}}

\address[7]{\orgdiv{School of Biomedical Engineering \& Imaging Sciences}, \orgname{King’s College London}, \orgaddress{\country{UK}}}

\corres{Hannah Eichhorn, \newline Institute of Machine Learning in Biomedical Imaging,  \newline Helmholtz Munich,  \newline Ingolstädter Landstr. 1, \newline 85764 Neuherberg, Germany. \newline\email{hannah.eichhorn@tum.de}}

% \corres{Prof. Julia A. Schnabel, \newline Institute of Machine Learning in Biomedical Imaging,  \newline Helmholtz Munich,  \newline Ingolstädter Landstr. 1, 85764 Neuherberg, Germany. \newline\email{julia.schnabel@tum.de}}

%%%%%%%%%%%%%%%%%%%%
% Abstract
%%%%%%%%%%%%%%%%%%%%
\abstract[Abstract]{
\section{Purpose} \tstar{} quantification from gradient echo magnetic resonance imaging is particularly affected by subject motion due to the high sensitivity to magnetic field inhomogeneities, which are influenced by motion and might cause signal loss. Thus, motion correction is crucial to obtain high-quality \tstar{} maps.
\section{Methods} We extend our previously introduced learning-based physics-informed motion correction method, PHIMO, by utilizing acquisition knowledge to enhance the reconstruction performance for challenging motion patterns and increase PHIMO's robustness to varying strengths of magnetic field inhomogeneities across the brain. We perform comprehensive evaluations regarding motion detection accuracy and image quality for data with simulated and real motion.
\section{Results} Our extended version of PHIMO outperforms the learning-based baseline methods both qualitatively and quantitatively with respect to line detection and image quality. Moreover, PHIMO performs on-par with a conventional state-of-the-art motion correction method for \tstar{} quantification from gradient echo MRI, which relies on redundant data acquisition.
\section{Conclusion} PHIMO's competitive motion correction performance, combined with a reduction in acquisition time by over 40\% compared to the state-of-the-art method, make it a promising solution for motion-robust \tstar{} quantification in research settings and clinical routine.
}


% \wordcount{\quickwordcount{main}}
% wordcount at submission: 5000

% \jnlcitation{}

\maketitle
%TC:endignore

%%%%%%%%%%%%%%%%%%%
% Introduction
%%%%%%%%%%%%%%%%%%%%
\section{Introduction}\label{sec:intro} 
Patient head motion during image acquisition remains a major challenge for brain magnetic resonance imaging~(MRI). \tstar{}-weighted gradient echo (GRE) acquisitions are particularly affected by motion due to their sensitivity to local magnetic field (B$_0$) inhomogeneities, which are influenced by head motion and can induce signal loss.\cite{Liu_2018,Marques_2005} 
The impact of motion-induced B$_0$ inhomogeneity changes on \tstar{}-weighted GRE images increases with increasing echo times.\cite{Magerkurth_2011} 
Since \tstar{} quantification requires multiple echoes, motion may lead to erroneous \tstar{} parameter maps, 
% Thus, motion might lead to erroneous  from multi-echo GRE MRI,
which can further affect derived parameter maps, such as the susceptibility related R2’ relaxation rate in the oxygenation-sensitive multi-parametric quantitative BOLD (mqBOLD) technique. \cite{Hirsch_2014,Eichhorn_2023_ISMRM} The current motion correction (MoCo) method for multi-echo GRE data, which is employed for mqBOLD MRI, relies on redundant acquisitions of the k-space center and thus, significantly increases the overall acquisition time.\cite{Noth_2014}

Various prospective and retrospective approaches have previously been proposed for rigid-body MoCo in brain MRI.\cite{Godenschweger_2016} A few methods also take into account motion-induced B$_0$ inhomogeneity changes. \cite{vanderKouwe_2006,Matakos_2010,Brackenier_2022,Hewlett_2024ismrm} However, retrospective methods do not rely on external hardware or sequence modifications for motion and field estimates, and are thus particularly challenged by the signal loss due to motion-induced B$_0$ inhomogeneity variations.\cite{Godenschweger_2016}

In recent years, deep learning has shown to be capable of learning complex patterns and dealing with inconsistent data and, thus, might offer promising solutions to address this challenge. 
Many learning-based MoCo approaches have already been proposed for brain MRI.\cite{Spieker_2023} MoCo can be addressed with image denoisers, using convolutional \cite{Pirkl_2022,Xu_2022} or generative adversarial networks. \cite{Johnson_2019,Kustner_2019,Oh_2021} Yet, relying exclusively on image data, these methods cannot guarantee consistency with the measured k-space data, which limits their clinical translation potential. Data consistency can only be ensured by integrating MoCo into the model- or learning-based image reconstruction process.\cite{Haskell_2019,Hossbach_2022,Singh_2023,Oksuz_2020} However, previous methods have been developed mainly for higher-resolution qualitative MRI or, if developed in the context of \tstar{} quantification, they do not enforce data consistency.\cite{Xu_2022} Furthermore, with respect to motion-induced B$_0$ inhomogeneity changes, Motyka et al.\cite{Motyka_2024} employ deep learning to predict motion-related B$_0$ inhomogeneity variations. Yet, to the best of our knowledge, no learning-based MoCo method exists that incorporates information on these B$_0$ variations into the reconstruction.

We have recently introduced PHIMO, a \textbf{PH}ysics-\textbf{I}nformed \textbf{M}otion c\textbf{O}rrection technique,\cite{Eichhorn_2024} which implicitly utilizes the above described B$_0$ inhomogeneity-induced signal loss. In particular, PHIMO employs a physics-informed loss function to detect motion-corrupted k-space lines and exclude them from a data consistent reconstruction.
This physics-informed loss has proven promising in detecting motion events, but the final image quality strongly depends on the performance of the reconstruction network, which is challenged by specific motion patterns, i.e., when central k-space lines need to be excluded. Additionally, varying strengths of B$_0$ inhomogeneities across the brain make a consistent estimation of exclusion masks challenging. 
In this work, we leverage MR acquisition physics to enhance the robustness of PHIMO with two major extensions, we validate the underlying assumptions and perform comprehensive evaluations. Our contributions are three-fold:
\begin{enumerate}
    \item We improve the performance of PHIMO's reconstruction module for scenarios where the subject moves during acquisition of the k-space center. 
    \item We accelerate the self-supervised line detection and improve PHIMO's robustness to varying strengths of inhomogeneities throughout the brain.
    \item We perform extensive evaluations and comparisons with other conventional and learning-based MoCo methods, using simulated and real motion data. 
\end{enumerate}

%%%%%%%%%%%%%%%%%%%%
% Theory
%%%%%%%%%%%%%%%%%%%%
\section{Theory}\label{sec:backgr}
\subsection{\tstar{} mapping from GRE MRI} \label{sec:backgr:t2star_mapping}
The effective transverse relaxation time \tstar{} can be determined from a series of GRE images $x = [x_1, ..., x_N]$ acquired at increasing echo times~$\mathtt{TE}_n$ for $n=1, ..., N$. Without motion and without B$_0$ inhomogeneities, the time evolution of the signal magnitude $s_n = \lvert x_n \rvert$ for an individual voxel can be approximated by:\cite{Magerkurth_2011} 
\begin{equation}
    s_n = s_0 \cdot \exp\left(-\frac{\mathtt{TE}_n}{T_2^*}\right)
    \label{eq:decay}
\end{equation}
with $s_0$ representing the signal magnitude at $\mathtt{TE}_n=0$. \tstar{}~and $s_0$ can thus be obtained by voxel-wise least-squares fitting of the multi-echo GRE data.
%

%
%TC:ignore
%%%%%% FIGURE - Overview %%%%%%
\begin{figure*}[t]
\centerline{\includegraphics[width=\linewidth]{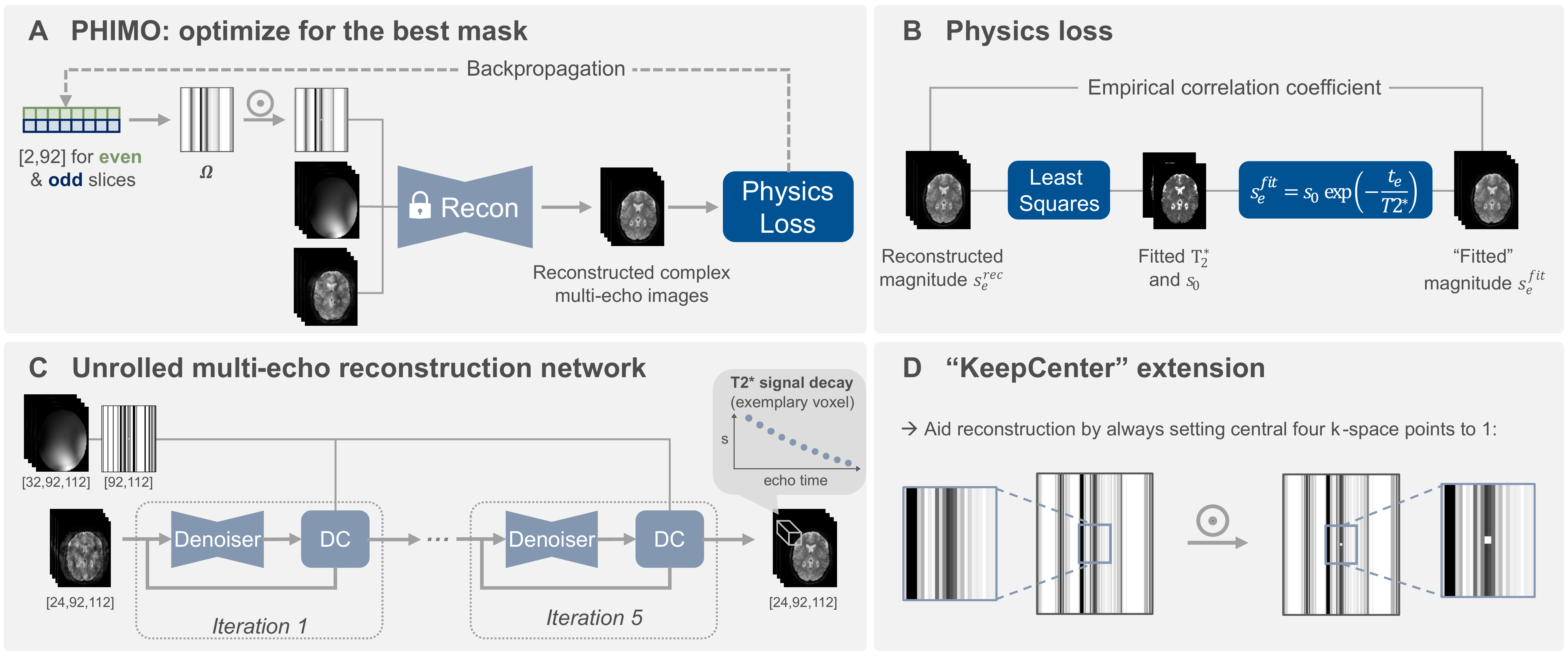}}
\caption{\ Illustration of PHIMO. (A) Subject-specific MoCo of \textit{motion-corrupted} data: optimizing one exclusion mask, $\Omega$, for even and one odd slices. 
(B) Empirical correlation coefficient as physics-informed loss for the self-supervised optimization of the exclusion masks in (A).
(C) Training an unrolled reconstruction network with randomly undersampled \textit{motion-free} multi-echo data.     
(D) \textit{KeepCenter} extension: always keep the central four k-space points regardless of their motion status to avoid severe contrast loss.
}\label{fig:PHIMO}
\end{figure*}
%TC:endignore
%
\subsection{MRI forward model with motion} \label{sec:backgr:forward_model}
The multicoil MRI forward model, accounting for motion, involves the sampling mask $\mathbf{S}_t$ (including the line-wise k-space acquisition pattern), the Fourier transform $\mathcal{F}$, the coil sensitivity maps $\mathbf{C}$, and the motion transform $\mathbf{U}_t$. To obtain the motion-affected k-space data $\hat{\mathbf{y}}$, these are applied to the motion-free image $x$ for each time point $t$:\cite{Atkinson_2023}
\begin{equation}
\label{eq:MRForward}
    \hat{\mathbf{y}} = \sum_{t=1}^{T} \mathbf{S}_t \mathcal{F} \mathbf{C} \mathbf{U}_t x
\end{equation}
For rigid-body motion in the presence of field inhomgeneities, $\mathbf{U}_t$ consists of translation and rotation transforms, $\mathbf{T}_t$ and $\mathbf{R}_t$, as well as a phase shift which is induced by the position-dependent in-plane B$_0$ inhomogeneities $\omega_t$ and increases with echo time $\mathtt{TE}_n$:\cite{Eichhorn_2023} %$\omega_{B_0}$
\begin{equation}
\label{eq:motion-transf}
    \mathbf{U}_t = e^{-2i\omega_t \mathtt{TE}_n} \mathbf{T}_t \mathbf{R}_t
\end{equation}

\subsection{Assumptions on motion patterns} \label{sec:backgr:kspace_center}
Head motion is commonly modeled as rigid-body motion that occurs at random timings, e.g. caused by the patient's discomfort or lack of attention. We assume that the patient is able to stay still for most of the scan with only occasional movements. Thus, it is possible to define an undersampling mask $\mathbf{\Omega}$ that excludes these individual motion events, so that the masked motion-corrupted data closely approximate the masked motion-free data $\mathbf{y}$:
\begin{equation}
    \mathbf{\Omega} \hat{\mathbf{y}} \simeq \mathbf{\Omega} \mathbf{y}
\label{eq:excl-motion}
\end{equation}

This allows us to split the MoCo into two subproblems: (1)~finding the optimal exclusion mask $\mathbf{\Omega}$  and (2)~reconstructing the undersampled k-space data~$\mathbf{\Omega} \hat{y}$. We have recently introduced a solution to these two subproblems,\cite{Eichhorn_2024} which we briefly describe together with the current major improvements in Section~\ref{sec:methods:phimo}.

If the subject moves during the acquisition of the k-space center, Eq.~\ref{eq:excl-motion} leads to the exclusion of central k-space lines, which contain important information about the overall image contrast. Completely disregarding such lines severely hampers the correct reconstruction of image intensities and thus, falsifies the subsequent \tstar{} quantification, depending of the extent of excluded data.. 
To minimize the impact of excluding central k-space lines, we propose to utilize the fact that 
the central k-space point $\mathbf{y}_{0,0}$ corresponds to the zero spatial frequency, representing the mean image intensity: $\mathbf{y}_{0,0} = \Bar{x}$.
Under the assumption that the mean image intensity does not significantly change for small amounts of motion ($\Bar{\hat{x}} \simeq \Bar{x}$), we set the central entry of the exclusion mask, $\mathbf{\Omega_{0.0}}$, to 1, independent of the central line's motion status. In practice, for an even matrix size we have to set the four central points in $\mathbf{\Omega}$ to 1.

%%%%%%%%%%%%%%%%%%%%
% Methods
%%%%%%%%%%%%%%%%%%%%
\section{Methods}\label{sec:methods}
\subsection{PHIMO} \label{sec:methods:phimo}
We have previously introduced a physics-informed MoCo method, which builds on Eq.~\ref{eq:excl-motion} and splits the MoCo into (1) a physics-informed detection of motion-corrupted k-space lines and (2) a reconstruction of undersampled data.\cite{Eichhorn_2024, Eichhorn_2024_ISMRM} Figure~\ref{fig:PHIMO} illustrates these steps. In the following, we provide a short overview of PHIMO and elaborate on the refinements we introduce in this extension. For the remainder of this article, we refer to the previous version of PHIMO \textit{without} the proposed refinements as "PHIMO-wo" and to the current version as "PHIMO".

\subsubsection{Unrolled multi-echo reconstruction} \label{sec:methods:phimo:recon}
For the reconstruction task, we utilize an unrolled network (Figure \ref{fig:PHIMO}C), which is trained on randomly undersampled \textit{motion-free} data, utilizing a mean squared error loss. The reconstruction alternates five times between a CNN-based denoiser and gradient descent data consistency layer, with independent weights for each iteration. Real and imaginary components of the input image and multiple echoes are stacked in the channel dimension. For more implementation details we refer to our previous publication \cite{Eichhorn_2024}. 

\subparagraph{Refinements:} As outlined in Section~\ref{sec:backgr:kspace_center}, we adapt our previous implementation to improve the reconstruction performance in scenarios where the subject moves during the acquisition of the k-space center. For this \textit{KeepCenter} setting, we set the four central pixels in the undersampling mask to 1, already during training of the reconstruction network (Figure \ref{fig:PHIMO}D).
For the comparison experiments (see Section~\ref{sec:methods:exp}), we train multiple networks with differently generated masks, e.g. with and without the \textit{KeepCenter} setting.

\subsubsection{Self-supervised k-space line detection} \label{sec:methods:phimo:moco}
For detecting affected k-space lines in \textit{motion-corrupted} data, PHIMO optimizes for the best exclusion mask with a self-supervised physics loss\cite{Eichhorn_2024} (Figure~\ref{fig:PHIMO}A and B). The physics loss is calculated as the empirical correlation coefficient between reconstructed signal intensities $s_e^{rec}$ and mono-exponentially “fitted” intensities $s_e^{fit}$ (Eq.~\ref{eq:decay}):
\begin{equation}
L_{phys} = 1 - \frac{\sum_e (s_e^{rec} - \overline{s}^{rec}) (s_e^{fit} - \overline{s}^{fit})}
    {\sqrt{\sum_e (s_e^{rec} - \overline{s}^{rec})^2 (s_e^{fit} - \overline{s}^{fit}})^2},
\end{equation}
with the mean over all echoes $\overline{s} = \frac{1}{E} \sum_{e} s_e$. $L_{phys} $ makes use of motion-related B$_0$ inhomogeneity changes, which disturb the mono-exponential signal evolution. \cite{Noth_2014} $L_{phys}$ is calculated voxel-wise across echoes and averaged within the brain mask.

\subparagraph{Refinements:} To reduce reconstruction times, we directly optimize a parameter vector, representing each slice's exclusion mask, instead of using a multilayer perceptron for predicting the exclusion mask.\cite{Eichhorn_2024} Furthermore, we incorporate knowledge of the interleaved MR acquisition scheme and  optimize only one exclusion mask for even and odd slices, respectively (Figure~\ref{fig:PHIMO}A), instead of individual masks for all slices. These \textit{Even/Odd} masks are optimized on the most inferior eight slices with mean susceptibility gradients $<80~\mu T/m$. 
This criterion selects slices with moderate susceptibility gradients, enabling a robust selection of hyperparameters that remains effective across different datasets without the need for fine-tuning.
The resulting masks are then applied to all slices during inference.  

PHIMO's output is a mask with continuous values between 0 and 1 whose size corresponds to the number of phase encoding (PE) lines. We do not binarize the output to benefit from smoother loss minimization. 
To achieve a stable optimization, we include a regularization term to enforce a minimal number of excluded lines:
\begin{equation}
     L_{reg} = (1- \frac{1}{Y}\sum_{y=1}^{Y} \Omega_{y}),
\end{equation}
with $Y$ representing the number of PE lines. 
In order to particularly penalize the exclusion of central k-space, we add a second term with stronger weight on the central ten k-space lines:
\begin{equation}
     L_{reg} = (1- \frac{1}{Y}\sum_{y=1}^{Y} \Omega_{y}) + 2 \cdot (1-\frac{1}{10} \sum_{y=Y_1}^{Y_2} \Omega_{y}), 
\end{equation}
with  $Y_1=\frac{Y}{2}-5$ and $Y_2=\frac{Y}{2}+5$.

\subparagraph{Training Details:}
PHIMO is optimized for 100 epochs with Adam,\cite{Adam_2015} a learning rate of 0.01, a batch of 20 slices and the regularization term is weighted with $\lambda = 0.005$. Differentiable least-squares fitting is implemented with \texttt{torch.linalg.lstsq}. The optimization takes less than three minutes per subject.

\subsection{Data acquisition} \label{sec:methods:data}
We have acquired multi-coil k-space data from 19 volunteers (27.0 ± 2.8 years, 7 females) on a 3T Philips Elition X MR scanner (Philips Healthcare, Best, The Netherlands). The acquisition consists of an interleaved multi-slice 2D GRE sequence that acquires even and odd slices in two packages
(12 echoes, 36 slices, 92~PE~lines, $\mathtt{TE}_1$=$\Delta\mathtt{TE}$=5~ms, TR=2300~ms, voxel size: 2$\times$2$\times$3~mm, 32-channel head coil). 
The study has been approved by the local ethics committee (approval numbers \mbox{440/18 S-AS}, \mbox{2023-386-S-SB}). 

We have performed repeated scans under two conditions, instructing the subject to (1) remain still, and (2) to move randomly (e.g. imitating sneezing or coughing). To evaluate PHIMO's motion detection performance, we have provided five subjects with precise timings on when to move. We refer to this as \textit{motion timing experiment}. To explore different motion types, we instructed one subject ("\textit{MoCo validation subject}") to perform four motion patterns with three different amplitudes each. 
%This subject is referred to as \textit{MoCo validation subject}. 
Moreover, we have collected half- and quarter-resolution data for all subjects under all motion conditions to compare PHIMO's performance with the state-of-the-art MoCo method \cite{Noth_2014}.

\subparagraph{Preprocessing:} \label{sec:methods:data:preproc}
The data are divided subject-wise into  train/validation sets for reconstruction (6/3 subjects) and validation/test sets for MoCo (1/9 subjects), including only slices with more than 20\% brain voxels. 
For input to the reconstruction network, data are normalized per 3D volume based on the maximum image magnitude.

\subsection{Motion simulation} \label{sec:methods:simulation}
To expand our evaluations and to train a supervised baseline method (Section~\ref{sec:methods:exp:comp}), we simulate head motion in motion-free multi-coil images. The simulation of rigid-body motion and random B$_0$ inhomogeneity variations follows our previously presented physics-aware simulation pipeline.\cite{Eichhorn_2023} The simulations are based on real head motion, i.e., 82 motion curves extracted from in-house fMRI time series.\footnote{Unpublished data from two ongoing studies with approval numbers
472/16 S and 15/20 S, independent cohorts from above imaging cohorts.}
These curves are divided into training, validation, and test sets (40/15/27 curves).

Principal component analysis is used to determine the largest modes of variations of
the training motion curves. The largest 20\% of all $N$ principal components, $pc_i(t)$, are combined with random weights $\alpha_i$
and added to the mean curve $\overline{T(t)}$ to generate 90 augmented training samples:~\cite{Cootes_1995}
\begin{equation}
    T_{train}(t) = \overline{T(t)} + \sum_{i=1}^{0.2\cdot N} \alpha_i \cdot pc_i(t).
\end{equation}

\subparagraph{Implementation details:} \label{sec:methods:simulation:details}
We simulate motion only for PE lines, where the average displacement of points within a sphere with radius 64~mm - modelling the patient's head - exceeds a threshold of 2~mm, the voxel size of our data. To circumvent the requirement of registration when calculating full-reference metrics of the simulated images relative to the original motion-free image, the motion curves are shifted so that the median motion state - measured by average displacement - is positioned at zero.

\subsection{Experiments} \label{sec:methods:exp}

\subsubsection{Validation of assumptions} \label{sec:methods:exp:validation}
PHIMO relies on assumptions for the reconstruction and line detection steps, including the proposed \textit{KeepCenter} and \textit{Even/Odd} extensions. To ensure the validity of these assumptions, we conduct experiments using the MoCo validation subject, i.e., the subject performing four different motion patterns with three different amplitudes.
Regarding the reconstruction choices, we test if the central four k-space points are affected by motion (Section~\ref{sec:backgr:kspace_center}). For this, we compare reconstructions of only the central four k-space points from acquisitions without and with intentional motion. Additionally, we investigate the impact of \textit{KeepCenter} on the reconstruction quality when excluding several central k-space lines. To assess this, we compare the mean absolute error (MAE) of \tstar{} maps obtained from networks trained without and with \textit{KeepCenter}. 

For the line detection choices, we first validate that the proposed physics-informed loss function correlates with the degree of motion that affects the GRE data. For this, we compare the loss value for motion-free images and three levels of intentional motion. Furthermore, we evaluate the effects of the \textit{Even/Odd} extension on the detection robustness across the brain by analyzing the line detection accuracy relative to the reference mask for different slice numbers.

\subsubsection{Comparative evaluation on test set} \label{sec:methods:exp:comp}
We compare the proposed refinements of PHIMO  to two learning-based alternatives, outlier-rejecting bootstrap aggregation (ORBA)\cite{Oh_2021} and supervised line detection (SLD)\cite{Oksuz_2020,Eichhorn_2023}, as well as to the previous version, \mbox{PHIMO-wo},\cite{Eichhorn_2024} and the state-of-the-art MoCo for \tstar{} quantification from GRE MRI with redundant k-space acquisition (HR/QR).\cite{Noth_2014}

\subparagraph{ORBA \cite{Oh_2021}} approaches motion as a probabilistic undersampling problem and averages reconstructions with 15 random bootstrap masks that each randomly exclude some motion events. For a fair comparison, we implement ORBA by training a separate unrolled reconstruction network with variable density masks at a fixed exclusion rate of 0.5, utilizing the \textit{KeepCenter} setting.

\subparagraph{SLD \cite{Oksuz_2020,Eichhorn_2023}} utilizes simulated data for training a convolutional neural network to predict exclusion masks for a given motion-corrupted k-space in a supervised fashion. The training of the line detection network follows the details in our previous publication \cite{Eichhorn_2023}, but we reconstruct the resulting undersampled data with the same reconstruction network as used for PHIMO.

\subparagraph{PHIMO-wo \cite{Eichhorn_2024}} does not use the \textit{KeepCenter} setting during training and inference of the reconstruction network and optimizes an individual mask for every slice. For a robust optimization PHIMO-wo requires a second regularization term that enforces a small variation of the masks for adjacent slices. For more details refer to Eichhorn et al.\cite{Eichhorn_2024}.

\subparagraph{HR/QR \cite{Noth_2014}} leverages additional half- and quarter-resolution acquisitions and calculates a weighted average of the three/two acquisitions of each PE line in the central half/quarter of k-space, in order to suppress individual motion events in one of the acquisitions. The reliance on redundant acquisition of the k-space center prolongs the acquisition time from 3~min~39~s to 6~min~25~s.

\subsection{Evaluation metrics} \label{sec:methods:eval}
\subsubsection{Line detection}  \label{sec:methods:line_det}
We evaluate PHIMO's line detection performance by calculating the MAE between predicted and reference masks, and the accuracy, quantified by the fraction of correctly classified lines when thresholding the predictions at a level 0.5. Additionally, we analyze the precision-recall-curve for various thresholds. 
Precision measures the fraction of lines that are correctly predicted as motion-corrupted, relative to the total number of lines predicted as motion-corrupted.
Recall measures the fraction of lines that are correctly predicted as motion-corrupted, relative to the total number of actually motion-corrupted lines.
For simulated data, reference masks are derived from the time points where motion was simulated (Section~\ref{sec:methods:simulation}). For real motion data, reference masks are only available for the motion timing experiment, where the motion timings can be converted into masks using the known k-space acquisition scheme.

\subsubsection{Image quality}  \label{sec:methods:img_qual}
We quantitatively evaluate the quality of the \tstar{} maps based on MAE, structural similarity (SSIM), feature similarity (FSIM) and perceptual image patch similarity (LPIPS), which have been shown to correlate with radiological evaluation of image quality.\cite{Marchetto_2024,Eichhorn_2025_ISMRM} All metrics are calculated per slice relative to the motion-free \tstar{} map for brain tissue voxels with susceptibility gradients smaller than $100~\mu T/m$. 

In the case of real motion (Section~\ref{sec:results:real}), we align  all acquisitions to the motion-free acquisition via 3D registration of the stacked slices using SPM12\cite{SPM12} \texttt{coregister}. To differentiate between the methods’ performances for more severe and minor motion, we categorize the nine test subjects based on visual assessment of the motion-corrupted images, resulting in five subjects with severe and four subjects with minor motion.

%
%TC:ignore
%%%%%% FIGURE - Validation %%%%%%
\begin{figure*}
\centerline{\includegraphics[width=0.95\linewidth]{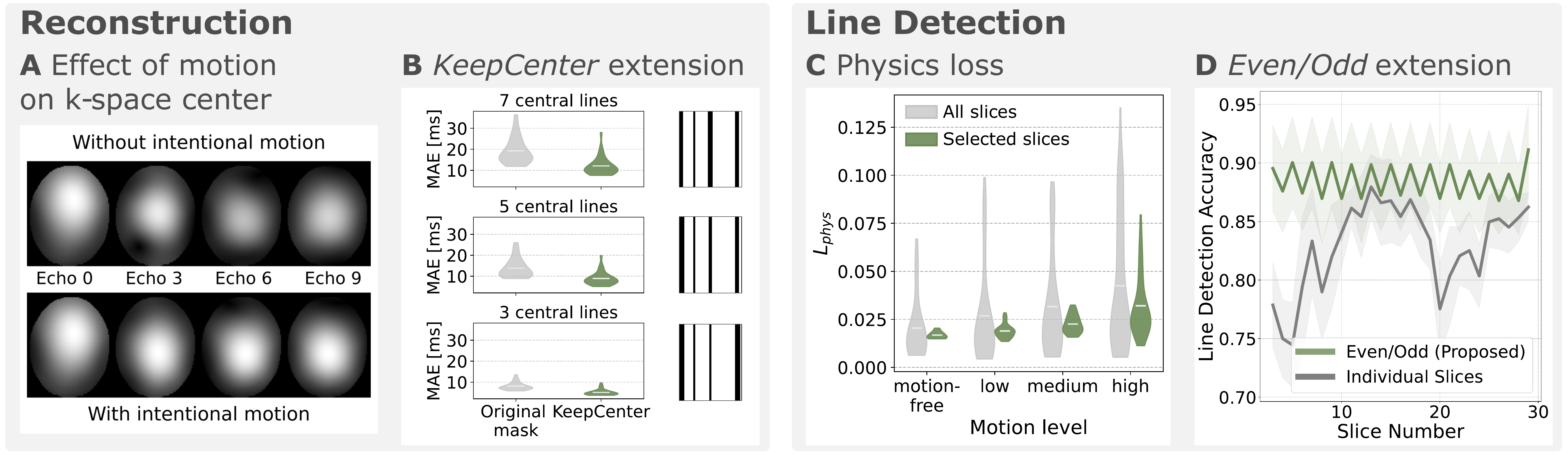}}
\caption{\ Investigating assumptions regarding reconstruction and line detection on the MoCo validation subject. 
(A)~Exemplary reconstructions of only four central k-space points for acquisitions without and with intentional motion (different echoes). 
(B)~MAE of \tstar{} maps resulting from the \textit{KeepCenter} and the original mask reconstructions for masks that exclude seven, five and three central k-space lines. Examples of reconstructed images can be found in the Supporting Information (Figure S1).
(C) Values of the physics loss for the motion-free acquisition compared to the acquisitions with low, medium and high motion amplitudes. The loss is once shown for all slices (gray) and once for only the inferior eight slices with mean susceptibility gradients $<80~\mu T/m$ (green), as we propose with the \textit{Even/Odd} extension.
(D) Slice-wise average line detection accuracy (relative to reference mask), comparing the \textit{Even/Odd} setting (green) to optimizing each slice individually (gray). Example exclusion masks are provided in the Supporting Information (Figure S2).
}\label{fig:Val}
\end{figure*}
%TC:endignore
%

\subsubsection{Statistical Analysis \& Implementation}  \label{sec:methods:stat_anal}
We use Wilcoxon signed rank tests and False-Discovery Rate correction for statistical testing. 
HR/QR motion correction and segmentation of anatomical scans are performed in MATLAB (R2022b) and SPM12 with custom programs.\cite{Kaczmarz_2020b} All other computations are performed in Python 3.8.12, using PyTorch 2.0.1 and MERLIN.\cite{HammernikKuestner2022} Our code is publicly available at \url{https://github.com/compai-lab/2025-mrm-eichhorn}.

%%%%%%%%%%%%%%%%%%%%
% Results
%%%%%%%%%%%%%%%%%%%%
\section{Results} \label{sec:results}
\subsection{Validation of assumptions} \label{sec:results:val}
We investigate the assumption that the proposed \textit{KeepCenter} setting improves PHIMO's performance for cases where the subject moves during the acquisition of the k-space center. Our analysis is based on the motion-free acquisition of the MoCo validation subject.
Although the exemplary reconstructed images from only four central k-space points, with and without intentional motion, exhibit some differences in details, their overall intensity distributions are comparable (Figure~\ref{fig:Val}A). 
Accordingly, for unrolled network reconstructions with masks that exclude several central k-space lines, the \textit{KeepCenter} extension reduces the MAE of the corresponding \tstar{} maps, compared to reconstructions with the original full masks (Figure~\ref{fig:Val}B). 
Examples of reconstructions for both settings are provided in the Supporting Information (Figure S1). These illustrate that the \textit{KeepCenter} extension avoids severe signal loss and recovers the contrast more correctly. 

We further test if our previously introduced physics loss\cite{Eichhorn_2024} reflects the degree of motion-corruption in the GRE images, using the repeated acquisitions of the MoCo validation subject. As illustrated in Figure~\ref{fig:Val}C, the value of the loss increases from motion-free images over low and medium to high motion amplitudes. Additionally, the larger variance of loss values for all slices compared to selected slices (based on susceptibility gradient strengths) confirms our assumption that the high variability of susceptibility strengths across the brain makes a robust optimization challenging.
Moreover, Figure~\ref{fig:Val}D demonstrates that the \textit{Even/Odd} extension, which is optimized for selected slices only, results in a robust line detection accuracy across slices, with only small differences between even and odd slices. In contrast, optimizing on individual slices leads to lower accuracies for inferior and superior slices.
This quantitative observation is also obvious in the example exclusion masks for different slices in the Supporting Information (Figure S2), which illustrates that the \textit{Even/Odd} extension leads to a more consistent line detection throughout the brain.
%
%TC:ignore
%%%%%% FIGURE - Examples Validation %%%%%%
\begin{figure*}
\centerline{\includegraphics[width=0.8\linewidth]{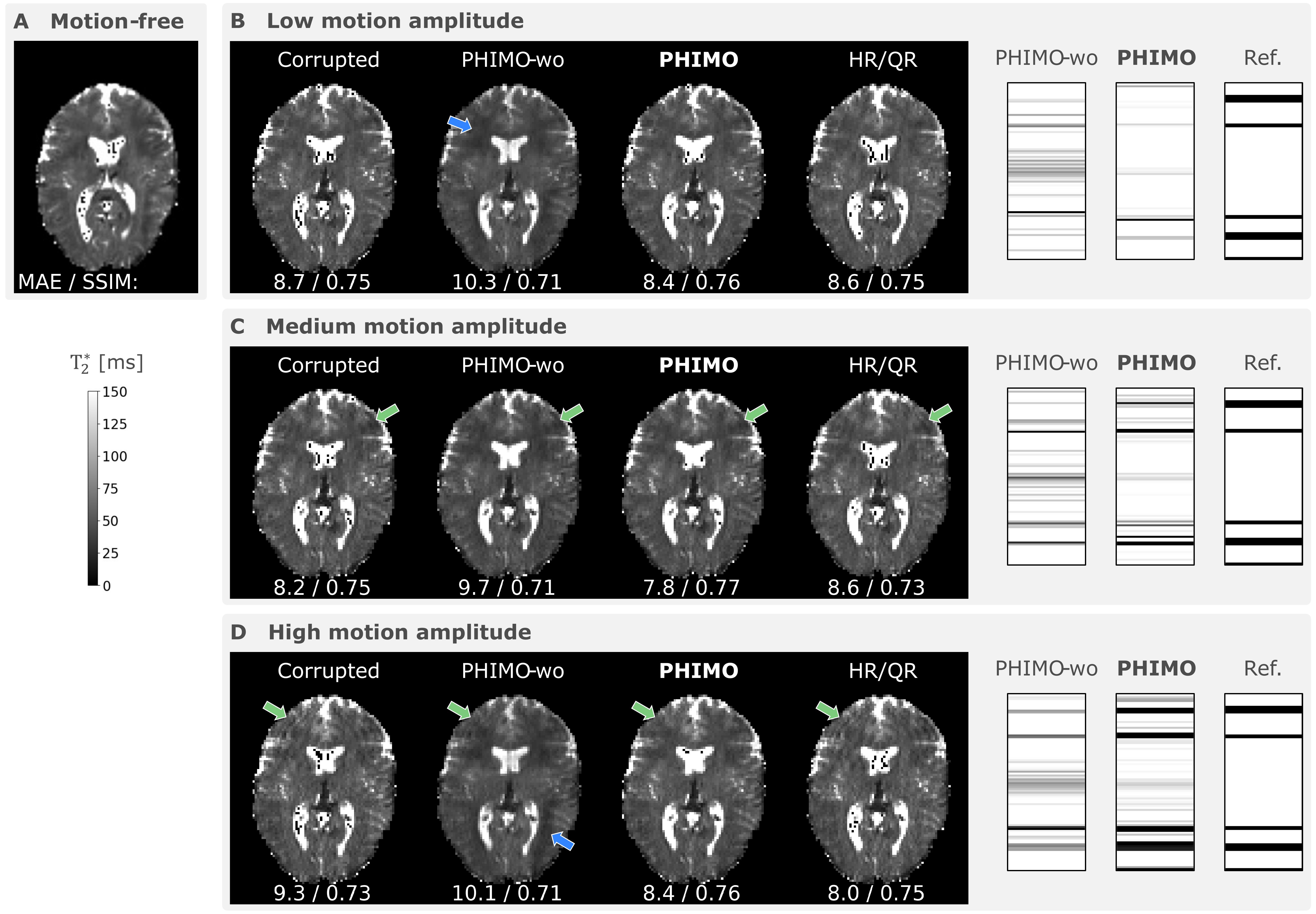}}
\caption{\ Testing PHIMO's adaption-ability to different motion levels for MoCo validation subject. The resulting \tstar{} map from PHIMO is compared to the uncorrected data, PHIMO-wo, HR/QR, and the separately acquired motion-free data (A).
The corresponding exclusion masks and the reference mask are presented on the right.
The three rows show different acquisitions with the same prescribed motion pattern but different motion amplitudes: low (B), medium (C) and high (D). 
Green arrows indicate motion artifacts that are best suppressed by PHIMO, blue arrows indicate visible \tstar{} estimation errors for PHIMO-wo.
MAE (in ms) and SSIM values for the visualized slices relative to the motion-free \tstar{} map are provided below each plot.
}\label{fig:Val_Exs}
\end{figure*}
%TC:endignore
%
%
%TC:ignore
%%%%%% FIGURE -Examples simulated motion %%%%%%
\begin{figure*}
\centerline{\includegraphics[width=\linewidth]{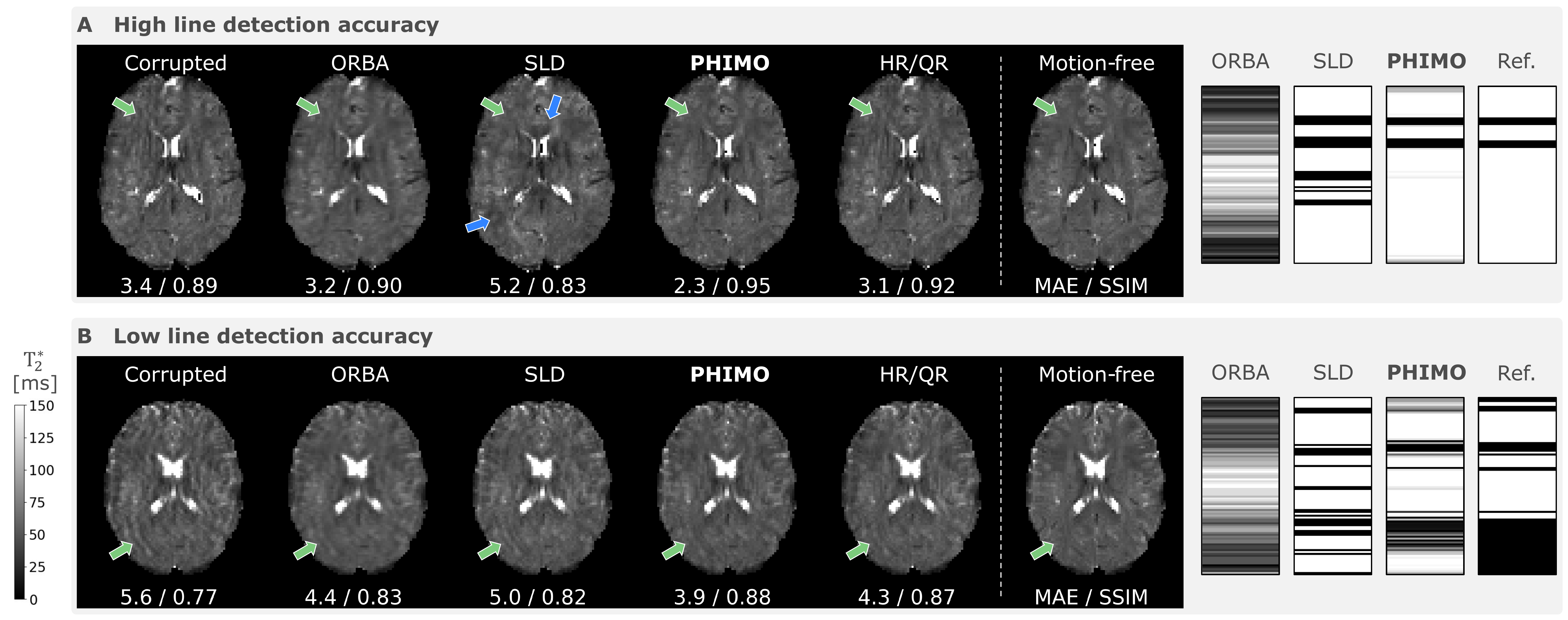}}
\caption{\ Qualitative examples for test subjects with simulated motion. The \tstar{} map resulting from PHIMO is compared to the uncorrected data (left), ORBA, SLD, HR/QR, and the separately acquired motion-free data (right).
The corresponding exclusion masks and the reference mask are additionally provided at the right.
The top row (A) shows data from one of the subjects with the highest line detection accuracies (0.989) and the bottom row (B) data from one of the subjects with the lowest accuracies for PHIMO (0.739). 
Green arrows indicate severe motion artifacts that are reduced by PHIMO and partially by HR/QR. Blue arrows indicate artifacts introduced by SLD.
MAE (in ms) and SSIM values for the visualized slices relative to the motion-free \tstar{} map are provided below each plot.}\label{fig:Ex_sim}
\end{figure*}
%TC:endignore
%
%
%TC:ignore
%%%%%% FIGURE - Line Detection simulated motion %%%%%%
\begin{figure}
\centerline{\includegraphics[width=0.9\linewidth]{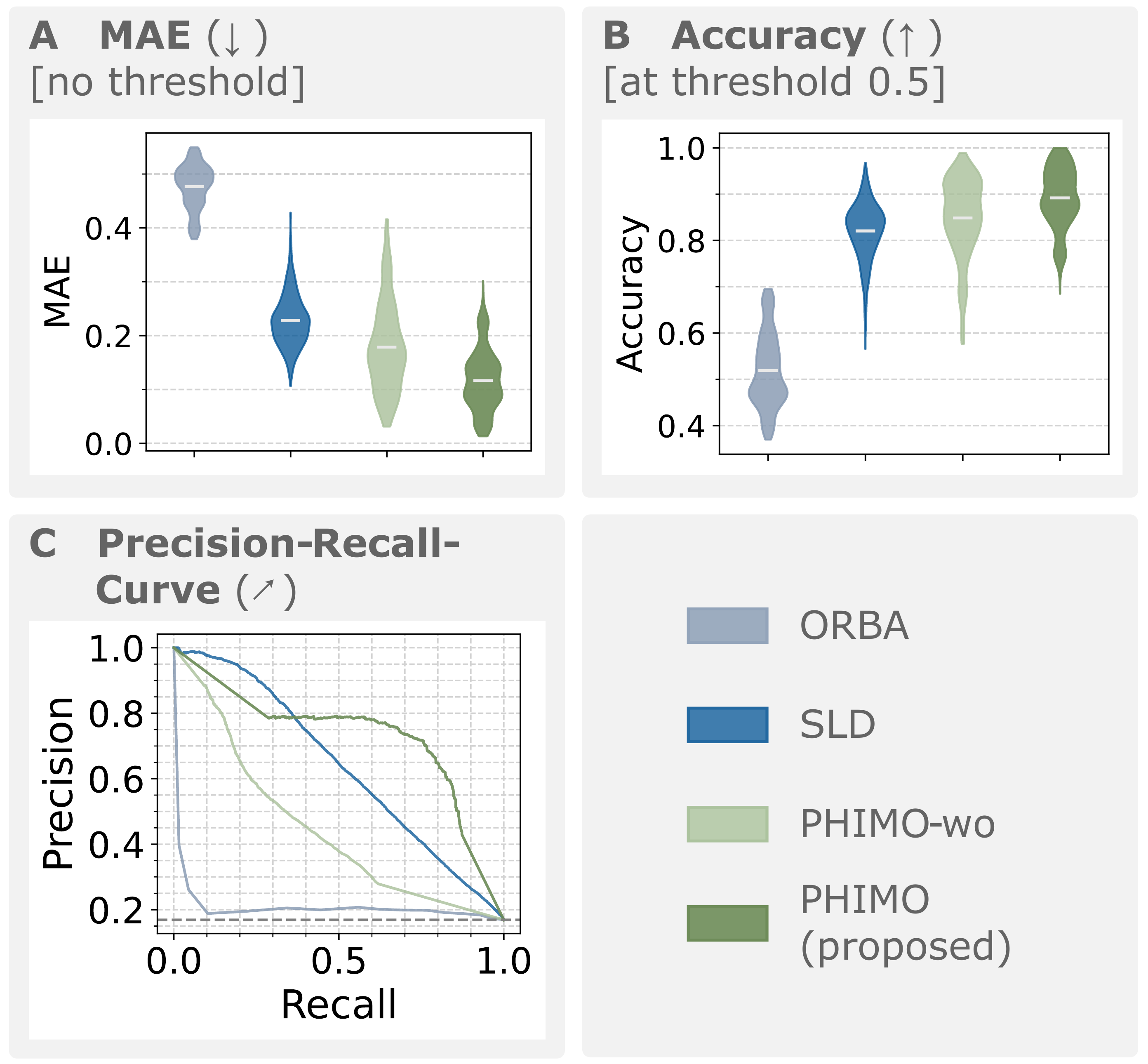}}
\caption{\ Line detection results for simulated test data. 
(A) MAE, (B) accuracy and (C) precision-recall-curve for all PE lines aggregated.
All comparisons in (A) and (B) are statistically significant (p$<$0.001).}\label{fig:LD_sim}
\end{figure}
%TC:endignore
%
%
%TC:ignore
%%%%%% FIGURE - IQMs simulated motion %%%%%%
\begin{figure}
\centerline{\includegraphics[width=0.9\linewidth]{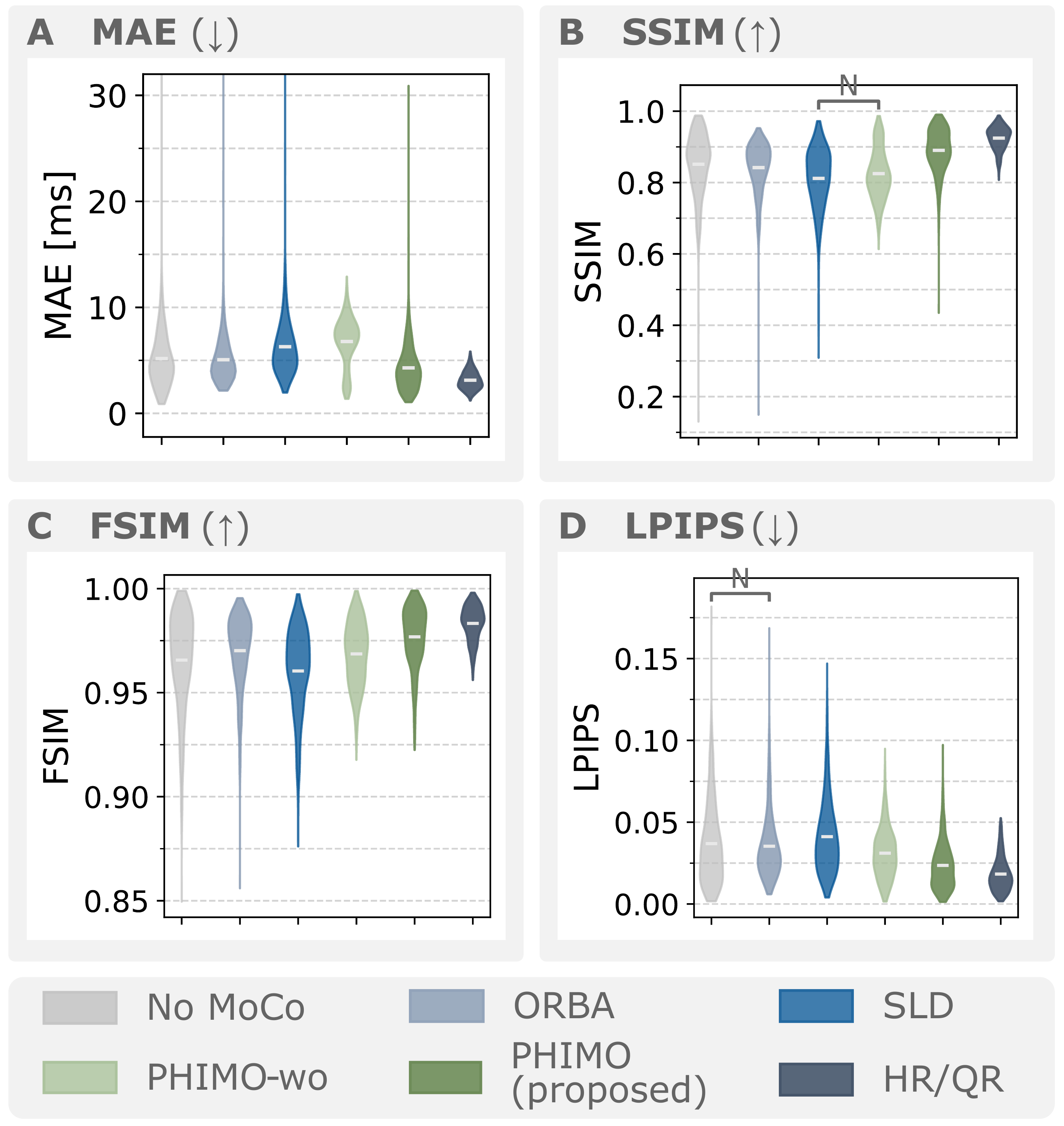}}
\caption{\ Quantitative evaluation of \tstar{} maps for test data with simulated motion. 
Image quality metrics (A) MAE, (B) SSIM, (C) FSIM and (D) LPIPS are calculated relative to the motion-free \tstar{} map. The proposed PHIMO improves compared to the uncorrected \tstar{} map and outperforms PHIMO-wo as well as the learning-based baselines. Gray brackets indicate comparisons with no statistical significance (p$>$0.001).
}\label{fig:IQMs_sim}
\end{figure}
% Outlier values for MAE: img_motion - MAE values larger than 32: [59.52427287]; img_orba - MAE values larger than 32: [62.9135801; img_sld - MAE values larger than 32: [39.49294632 35.87250271]
%TC:endignore
%

Additionally, we use the repeated acquisitions of the MoCo validation subject to compare PHIMO to PHIMO-wo and test if PHIMO is able to adapt the predicted masks to varying levels of motion corruption. Figure~\ref{fig:Val_Exs} shows examples for three repeated acquisitions with different motion amplitudes but approximately the same pattern. The mask examples validate that PHIMO is able to adapt the predicted masks to different motion levels. PHIMO's predictions match the reference more closely than PHIMO-wo. Furthermore, PHIMO recovers the \tstar{} values more accurately than PHIMO-wo and outperforms HR/QR in removing ringing artifacts for this motion pattern. Due to space limitations, we omit examples for PHIMO-wo below (Sections~\ref{sec:results:simulated} and \ref{sec:results:real}), but include it in the quantitative evaluations.

%%%%%%%%%%%%%%%%%%%% Simulated Data %%%%%%%%%%%%%%%%%%%%
\subsection{Evaluation on simulated test data}  \label{sec:results:simulated}
We compare example \tstar{} maps from different reconstructions of data without and with simulated motion (Figure~\ref{fig:Ex_sim}). PHIMO significantly reduces motion artifacts which mostly appear as ringing or wave-like patterns. The top row shows an example where the exclusion mask estimated by PHIMO closely resembles the prescribed reference. In the bottom row, where PHIMO does not detect all lines, the corresponding exclusion mask yet generally matches the pattern of the reference mask, which confirms that PHIMO can detect the dominant motion events. 
HR/QR also reduces severe motion artifacts, but shows more residual wave-like artifacts than PHIMO, particularly for the subject in the top row. The \tstar{} maps resulting from ORBA show an increased blurring but no clear reduction of the wave-like artifacts. SLD partially reduces the wave-like motion artifacts but introduces additional artifacts in some regions. The exclusion masks from SLD approximate the patterns of the reference masks but particularly overestimate motion events in the k-space center, especially for the subject in the top row, and miss more lines than PHIMO for the subject in the bottom row.
 
To analyze PHIMO's performance quantitatively, we compare PHIMO's line detection performance to PHIMO-wo, ORBA and SLD on the test data with simulated motion in Figure~\ref{fig:LD_sim}. PHIMO outperforms all comparison methods in terms of MAE and accuracy as well as in the precision-recall-curve.
Similarly, the quantitative evaluation of the resulting \tstar{} maps in Figure~\ref{fig:IQMs_sim} demonstrates that our proposed extension of PHIMO outperforms PHIMO-wo as well as the learning-based baselines, ORBA and SLD, and approaches the performance of HR/QR.

%%%%%%%%%%%%%%%%%%%% Real Motion Data %%%%%%%%%%%%%%%%%%%%
\subsection{Evaluation on real motion test data}  \label{sec:results:real}
%
%TC:ignore
%%%%%% FIGURE -Examples real motion %%%%%%
\begin{figure*}
\centerline{\includegraphics[width=\linewidth]{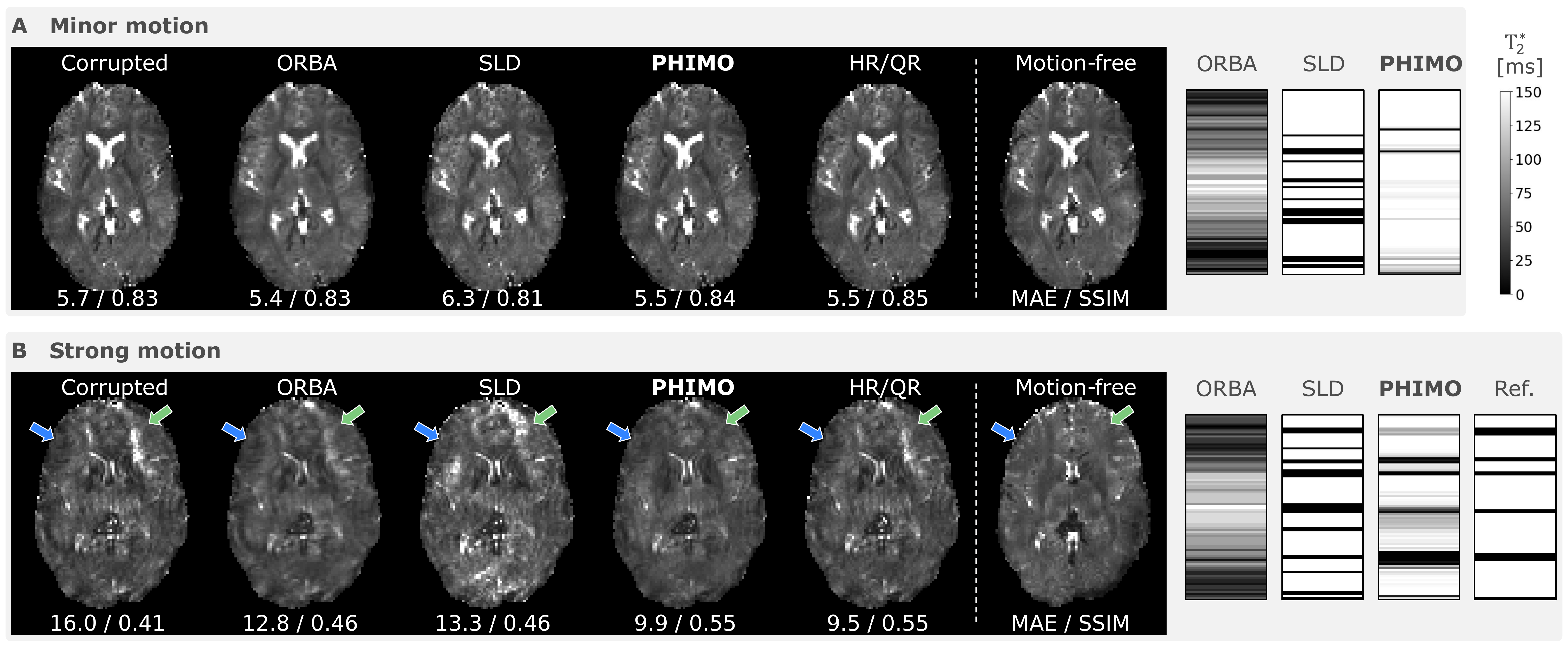}}
\caption{\ Qualitative examples for test data with real motion. The resulting \tstar{} map from PHIMO is compared to the uncorrected data (left), ORBA, SLD, HR/QR, and the separately acquired motion-free data (right).
%ORBA, SLD, HR/QR, the uncorrected and the motion-free data.
We additionally provide the corresponding exclusion masks and the reference mask (if available).
The top row (A) shows one subject with minor motion (no instructions on timing given) and the bottom row (B) one subject with strong motion (motion timing experiment). 
Green arrows indicate severe motion artifacts that are suppressed by PHIMO and partially by HR/QR, blue arrows indicate missing details for all methods.
MAE (in ms) and SSIM values for the visualized slices relative to the motion-free \tstar{} map are provided below each plot.}\label{fig:Ex_real}
\end{figure*}
%TC:endignore
%
%%% Qualitative results
We show example \tstar{} maps from test subjects with real motion in Figure~\ref{fig:Ex_real}.
For the subject exhibiting strong motion, PHIMO outperforms all comparison methods and effectively suppresses severe motion artifacts that otherwise cause regional \tstar{} misestimation errors. However, similar to all other methods, PHIMO is unable to fully recover all fine details compared to the separately acquired motion-free \tstar{} map for this level of head motion. The estimated exclusion mask matches the pattern of the reference mask (extracted from the prescribed timings), which confirms PHIMO's ability to detect real motion events. Similar to the simulated motion examples, SLD somewhat overestimates the exclusion mask, particularly in the k-space center.
For the subject exhibiting minor motion, the corrupted \tstar{} map shows only subtle motion artifacts and all methods either preserve its quality or lead to minor improvements in MAE and SSIM that are difficult to detect visually.
For this subject no reference mask is available, which precludes a direct evaluation of the predicted exclusion masks. However, PHIMO detects only a few motion-corrupted lines, as expected for such minor motion artifacts, while SLD excludes a large number of lines, especially in the k-space center.

%%% Quantitative results
%
%TC:ignore
%%%%%% FIGURE - Line Detection real motion %%%%%%
\begin{figure}
\centerline{\includegraphics[width=0.9\linewidth]{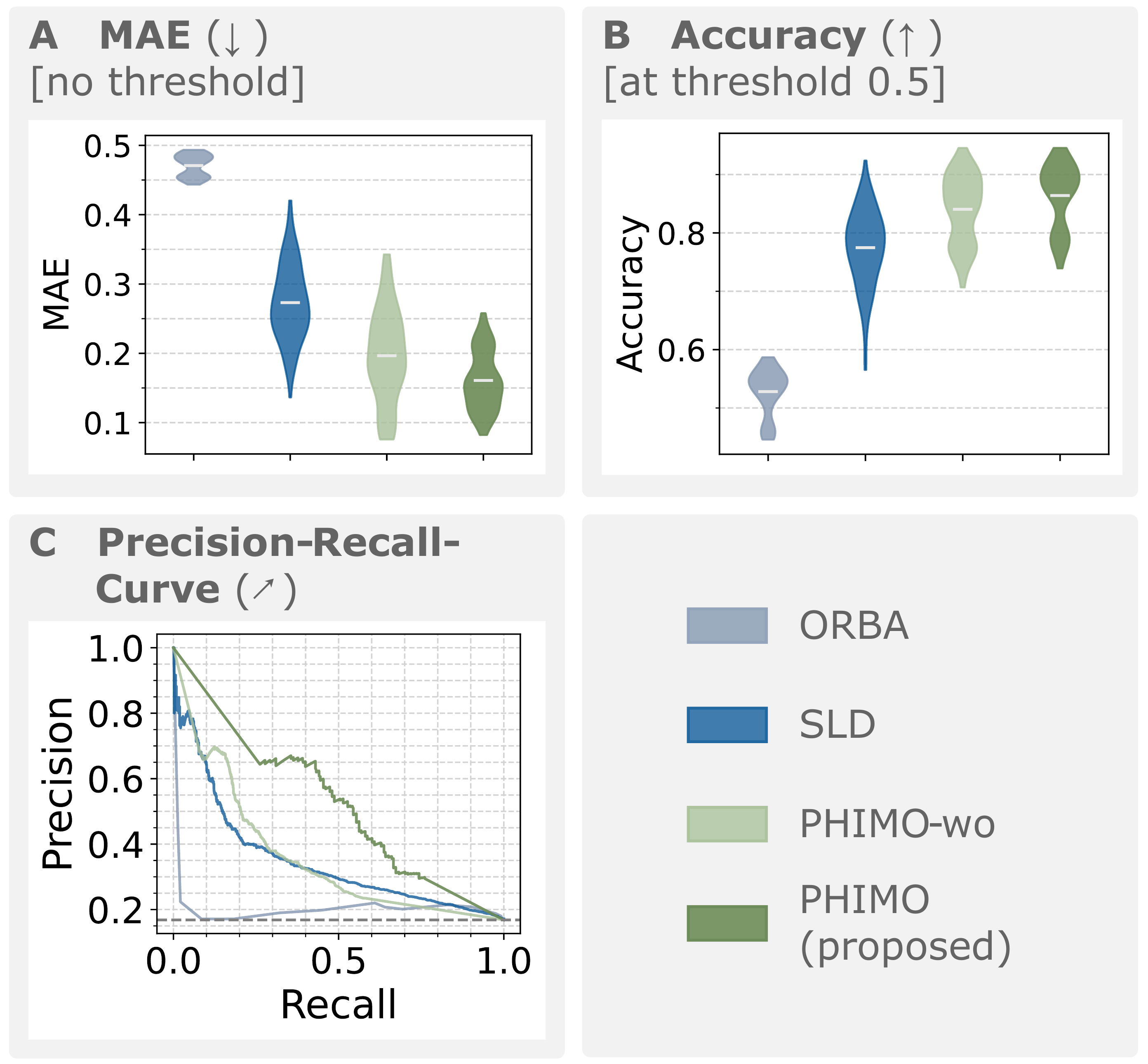}}
\caption{\ Line detection performance on real motion test data for the four subjects of the motion timing experiment. 
(A)~MAE, (B) accuracy and (C) precision-recall-curve.
Reference masks were generated from the prescribed motion timings using the known k-space acquisition scheme.
All comparisons in (A) and (B) are statistically significant (p$<$0.001).}\label{fig:LD_real}
\end{figure}
%TC:endignore
%
%
%TC:ignore
%%%%%% FIGURE - IQMs real motion %%%%%%
\begin{figure*}
\centerline{\includegraphics[width=\linewidth]{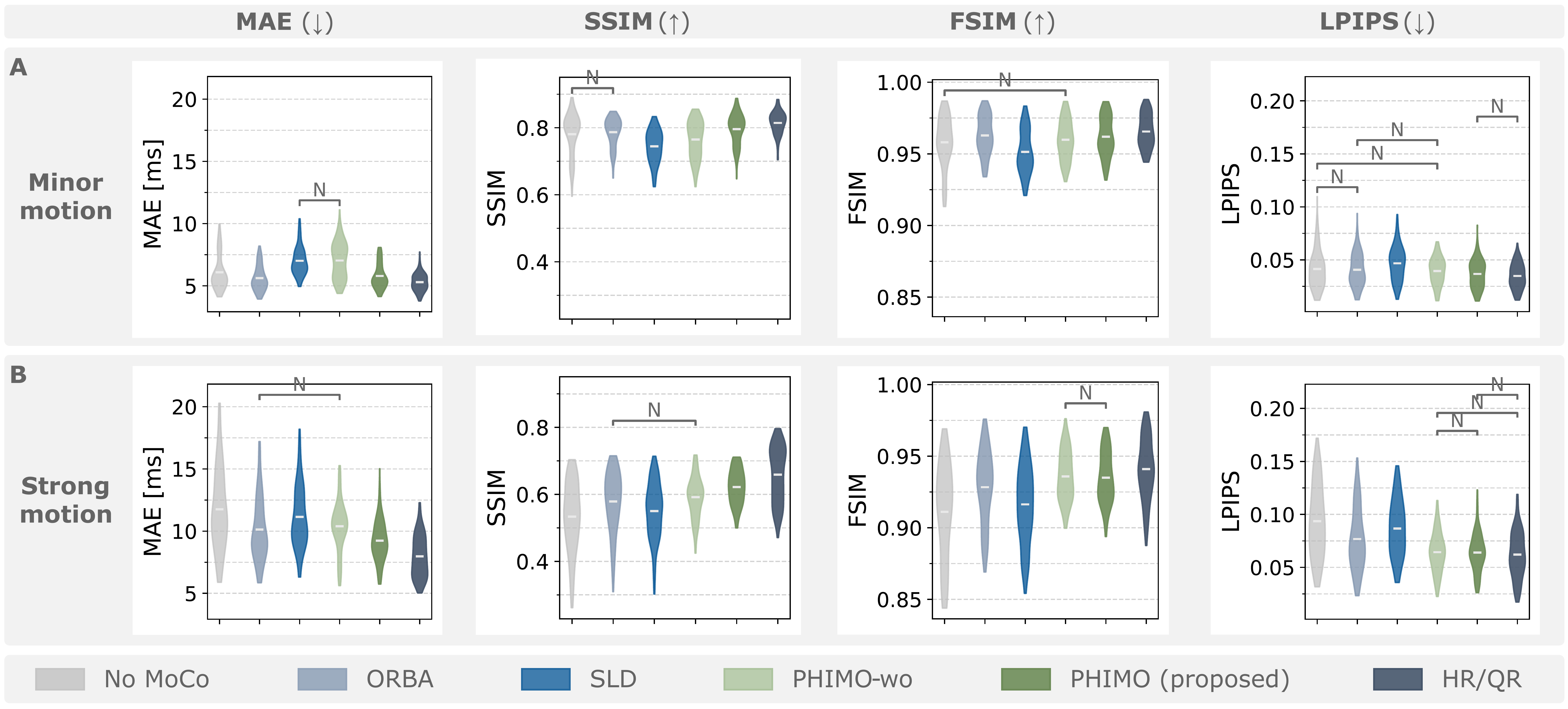}}
\caption{\ Quantitative evaluation of \tstar{} maps for real motion test data. 
Image quality metrics MAE, SSIM, FSIM and LPIPS relative to the motion-free \tstar{} map for four subjects with minor motion (A) and five subjects with strong motion (B). The proposed PHIMO improves compared to the uncorrected \tstar{} map and outperforms PHIMO-wo as well as the learning-based baselines (ORBA and SLD). Gray brackets indicate comparisons with no statistical significance (p$>$0.001).}\label{fig:IQMs_real}
\end{figure*}
%TC:endignore
%
Figure~\ref{fig:LD_real} quantitatively compares PHIMO's line detection performance on real motion data for the four test subjects that performed the motion timing experiment. In line with the qualitative observations from Figure~\ref{fig:Ex_real}, PHIMO shows superior performance compared to all other methods based on MAE, accuracy and the precision-recall-curve.

The quantitative evaluation of the \tstar{} maps based on different image quality metrics in Figure~\ref{fig:IQMs_real} supports our qualitative observations. For the subjects with strong motion, PHIMO improves all image quality metrics compared to the uncorrected maps and the learning-based baseline methods. The proposed extension of PHIMO outperforms PHIMO-wo in terms of MAE and SSIM and approaches the performance of HR/QR. Minor motion data generally show better metric values with small but consistent improvements by PHIMO and HR/QR. SLD, and to some extent \mbox{PHIMO-wo}, rather degrade image quality for minor motion, particularly in terms of MAE. A potential reason for this could be that they overestimate the amount of motion in the \mbox{k-space} center. For the subjects with minor motion, the average value of the exclusion mask for the central 10 lines is 0.95 for PHIMO, 0.84 for PHIMO-wo and 0.65 for SLD.

We additionally evaluate how PHIMO and the comparison methods deal with acquisitions without intentional motion. Example \tstar{} maps for their applications to apparently motion-free data are provided in the Supporting Information (Figure S3). In these examples, PHIMO and HR/QR preserve the quality of the motion-free data, whereas ORBA introduces blurring and SLD excludes many central k-space lines, partially resulting in severe artifacts.
This observation is also reflected in the summary of predicted exclusion masks by ORBA, SLD and PHIMO for the motion-free data in Table~\ref{tab:metrics_comparison}. Assuming that the cooperative volunteers remained still for the majority of the acquisitions, PHIMO's predicted exclusion masks, with an average of 1.3\% excluded lines, appear more realistic compared to SLD's average of 13.3\%.

%TC:ignore
\begin{table}[t]
\caption{\ Summary of exclusion masks predicted for motion-free data by ORBA, SLD and PHIMO.}
\centering
% \begin{tabular}{@{}lccc@{}}
\begin{tabular}{@{}p{2.5cm}p{1.5cm}p{1.5cm}p{1.5cm}@{}}
\toprule
\textbf{} & \textbf{ORBA} & \textbf{SLD} & \textbf{PHIMO} \\ \midrule
% \textbf{Average value of masks}          & $0.503 \pm 0.002$ & $0.867 \pm 0.013$ & $0.967 \pm 0.025$ \\
\raggedright \textbf{Average value of masks}          & $0.503$ & $0.867$ & $0.967$ \\
% \textbf{Value range}         & [0.501, 0.507] & [0.851, 0.887] & [0.908, 0.995] \\ 
% \midrule
% \textbf{Fraction of excluded lines} & $52.05 \pm 2.92$ & $13.30 \pm 1.32$ & $1.27 \pm 2.76$ \\
\raggedright \textbf{Fraction of excluded lines} & $52.1 \%$ & $13.3\%$ & $1.3\%$ \\
% \textbf{Value range: } & [48.91, 57.61] & [11.27, 14.92] & [0.00, 8.70] \\
\bottomrule
\end{tabular}
\begin{tablenotes}
\item For calculating the fraction of excluded lines, the masks were thresholded at 0.5.
\end{tablenotes}
\label{tab:metrics_comparison}
\end{table}
%TC:endignore

%%%%%%%%%%%%%%%%%%%%
% Discussion
%%%%%%%%%%%%%%%%%%%%
\section{Discussion}\label{sec:disc}
% Summary
We have proposed two major extensions to our previously introduced self-supervised and physics-informed learning-based MoCo method.\cite{Eichhorn_2024} This extended version of PHIMO improves the robustness of the self-supervised line detection by optimizing only one mask for even and odd slices, respectively. Additionally, the \textit{KeepCenter} extension improves the undersampled reconstruction and reduces \tstar{} quantification errors for challenging motion patterns affecting the k-space center. We have validated these extensions and their underlying assumptions with multiple validation experiments. Moreover, our comprehensive evaluations using simulated and real motion data have confirmed PHIMO's superiority to learning-based baseline methods. Our evaluations have further confirmed PHIMO's on-par performance with the state-of-the-art HR/QR MoCo method for \tstar{} quantification from GRE MRI,\cite{Noth_2014} while substantially reducing the acquisition time, since PHIMO does not rely on redundant acquisitions of the k-space center.

\subsection{PHIMO vs. PHIMO-wo}
Our validation experiments have supported our assumption that the reconstruction benefits from keeping the central four k-space points regardless of their motion status (Figure~\ref{fig:Val}). The information in the k-space center appears to not be severely affected by motion but aids the correct contrast reconstruction and subsequent \tstar{} quantification. 
Furthermore, we have demonstrated that the proposed physics loss - based on the \tstar{} decay fit - is sensitive to the level of motion-corruption. Moreover, the loss values are more tightly clustered when calculated on a selected subset of slices (based on susceptibility gradient strengths), which confirms our assumption that optimizing the exclusion mask only on a subset of slices allows for a more robust optimization. Together with our \textit{Even/Odd} extension, which groups slices that are acquired within a sufficiently narrow time frame, we achieve a more consistent detection performance throughout the brain.
Moreover, the visual and quantitative comparisons of PHIMO and PHIMO-wo\cite{Eichhorn_2024} demonstrate that the \textit{KeepCenter} and \textit{Even/Odd} extensions significantly improve the overall MoCo performance (Figure~\ref{fig:Val_Exs}).

\subsection{PHIMO vs. learning-based baselines}
We have compared PHIMO to two learning-based baseline methods, ORBA\cite{Oh_2021} and SLD.\cite{Oksuz_2020,Eichhorn_2023} These methods are also based on excluding motion-corrupted k-space lines and, thus, are applicable to \tstar{}-weighted GRE MRI with  B$_0$ inhomogeneities as secondary motion effects. For both simulated and real motion, ORBA results in blurry \tstar{} maps and cannot correct the observed motion artifacts, in particular wave-like patterns and ringing artifacts. This is not surprising, since - in contrast to PHIMO - ORBA does not specifically exclude motion-corrupted lines but averages the outcomes of several random masks. Our results demonstrate that PHIMO's explicit self-supervised line detection is superior to this probabilistic aggregation of random masks.

The second baseline method, SLD, is trained to exclude specific k-space lines in a supervised fashion on simulated data. However, compared to PHIMO, the predicted masks for unseen data are less accurate across all metrics, especially for real motion. We have observed that SLD tends to exclude too many lines in the k-space center, which particularly challenges the reconstruction and needs to be avoided in the absence of actual motion. In particular for motion-free data, SLD has partially introduced severe artifacts, while PHIMO has preserved the high quality of the motion-free maps~(Figure~S3). 
This overestimation of motion in the k-space center certainly contributes to SLD's lower image quality metrics, where PHIMO outperforms SLD for simulated and real motion data. PHIMO's superiority to SLD confirms the benefit of self-supervised scan-specific optimization, which avoids some generalization issues of supervised training with limited or potentially not fully realistically simulated training data.

\subsection{PHIMO vs. state-of-the-art HR/QR}
We have additionally compared PHIMO to HR/QR, which samples the k-space center three times, combining the motion-corrupted acquisition with additional half- and quarter-resolution acquisitions. 
Quantitatively, HR/QR achieves slightly higher values for most image quality metrics in simulated and real motion data; but PHIMO more closely approaches HR/QR's performance than any other comparison method. 
HR/QR's superior metric values must be considered in the context that it combines three acquisitions into a weighted average, which inherently improves SNR, even in the absence of motion, at the cost of longer acquisition times.
However, when the subject moves during the acquisition of peripheral k-space lines, HR/QR cannot correct the corresponding artifacts, since it only reacquires the k-space center. The examples in Figures~\ref{fig:Val_Exs},~\ref{fig:Ex_sim}~and~\ref{fig:Ex_real} demonstrate PHIMO's superiority for these scenarios.
Considering quantitative and qualitative results, PHIMO performs on-par with HR/QR, while significantly reducing the overall scan time by over 40\%, which is crucial for clinical translation.

\subsection{Limitations}
Our work is not without limitations. 
First, similar to all comparison methods, PHIMO assumes that the subject remains mostly still, with only occasional random movements. Thus, its performance depends on the specific motion patterns. In the Supporting Information (Figure S4), we provide an example that was excluded from the main analysis due to exaggerated motion in the k-space center, which challenges the undersampled reconstruction even with the \textit{KeepCenter} extension.
To improve PHIMO’s resilience to such severe motion patterns, one potential solution could involve randomizing the acquisition order in the PE direction, as proposed in the DISORDER framework,\cite{CorderoGrande_2020} and thus, distribute potentially long motion events across the entire k-space. 
Second, in the clinical routine short reconstruction times are essential. PHIMO currently requires up to three minutes for the scan-specific optimization of exclusion masks and final reconstruction. While this is already practical in research settings, where reconstructions can be performed offline, additional efficiency enhancements will be crucial to further reduce reconstruction times for clinical applications.
Third, by excluding motion-corrupted lines, PHIMO corrects motion-related B$_0$ inhomogeneity changes, but static B$_0$ inhomogeneities are still present and need to be addressed in a separate post-processing step.\cite{Hirsch_2013} In order to disentangle motion and static B$_0$ inhomogeneity artifacts in our analysis, we only evaluated the \tstar{}~maps for voxels with susceptibility gradients smaller than~$100~\mu T/m$.

%%%%%%%%%%%%%%%%%%%%
% Conclusion
%%%%%%%%%%%%%%%%%%%%
\section{Conclusion}\label{sec:concl}
In conclusion, we have developed two key extensions to our previously introduced physics-informed motion correction method, PHIMO. These extensions enhance PHIMO's robustness to highly variable susceptibility gradients across the brain and enhance its reconstruction performance for challenging motion patterns, in particular motion events occurring in the k-space center. Our comprehensive evaluation convincingly demonstrates that PHIMO outperforms two learning-based baseline methods and matches the performance of a state-of-the-art MoCo technique, while significantly reducing the total acquisition time. This makes PHIMO a promising option for motion-robust \tstar{} quantification from GRE MRI in clinical and research applications.

%TC:ignore

\section*{Acknowledgments}
Hannah Eichhorn and Veronika Spieker are partially supported by the Helmholtz Association under the joint research school ”Munich School for Data Science - MUDS”.

\subsection*{Conflict of interest}
Kilian Weiss is an employee of Philips GmbH Market DACH.

\subsection*{Data availability statement}
Our code is publicly available at \url{https://github.com/compai-lab/2025-mrm-eichhorn}. The data will be published once the manuscript is accepted.

\bibliography{references}%

\section*{Supporting information}
The following supporting information is available as part of the online article:

\vskip\baselineskip\noindent
\textbf{Figure S1.}
{Examples of reconstructed images (first echo) for networks trained conventionally (with original masks and no \textit{KeepCenter}) and with the \textit{KeepCenter} extension, compared to the fully-sampled image. From top to bottom the masks exclude three, five and seven central k-space lines. Green arrows indicate brain areas with more correctly recovered contrast in the \textit{KeepCenter} reconstructions.}

\noindent
\textbf{Figure S2.}
{Example exclusion masks estimated from inferior, middle and superior slices, comparing the \textit{Even/Odd} extension to an optimization of individual slices for two different motion patterns of the MoCo validation subject. For these acquisitions, the slice range based on the susceptibility gradient strength for the \textit{Even/Odd} optimization is [10, 17].
These examples demonstrate a more stable line detection profile across the brain for the \textit{Even/Odd} optimization compared to masks optimized for individual slices. }

\noindent
\textbf{Figure S3.}
{Qualitative examples for applying PHIMO and the comparison methods to apparently motion-free data. The resulting \tstar{} maps and, if available, exclusion masks are compared for the original data without intentional motion, PHIMO, ORBA, SLD and HR/QR for two different subjects.
Blue arrows indicate \tstar{} quantification errors introduced by SLD.}

\noindent
\textbf{Figure S4.}
{Acquisition excluded from the main analysis due to excessive motion in the k-space center (10~s / nine lines). 
The \tstar{} maps are compared for PHIMO, ORBA, SLD and HR/QR to the uncorrected and motion-free acquisitions. The respective exclusion masks are shown on the right.
Green arrows indicate areas where PHIMO, and to some extent HR/QR, clearly mitigate the extent of wave-like motion artifacts, blue arrows indicate missing details across all methods. 
PHIMO overestimates the exclusion mask, likely due to excessive motion in the k-space center, which challenges the reconstruction network even with the \textit{KeepCenter} extension. Note that the current standard, HR/QR, is also challenged by such an extreme motion case, which in clinical applications may ultimately require reacquisition.}

%TC:endignore

\appendix

\clearpage

\includepdf[pages=-, pagecommand={\thispagestyle{empty}}]{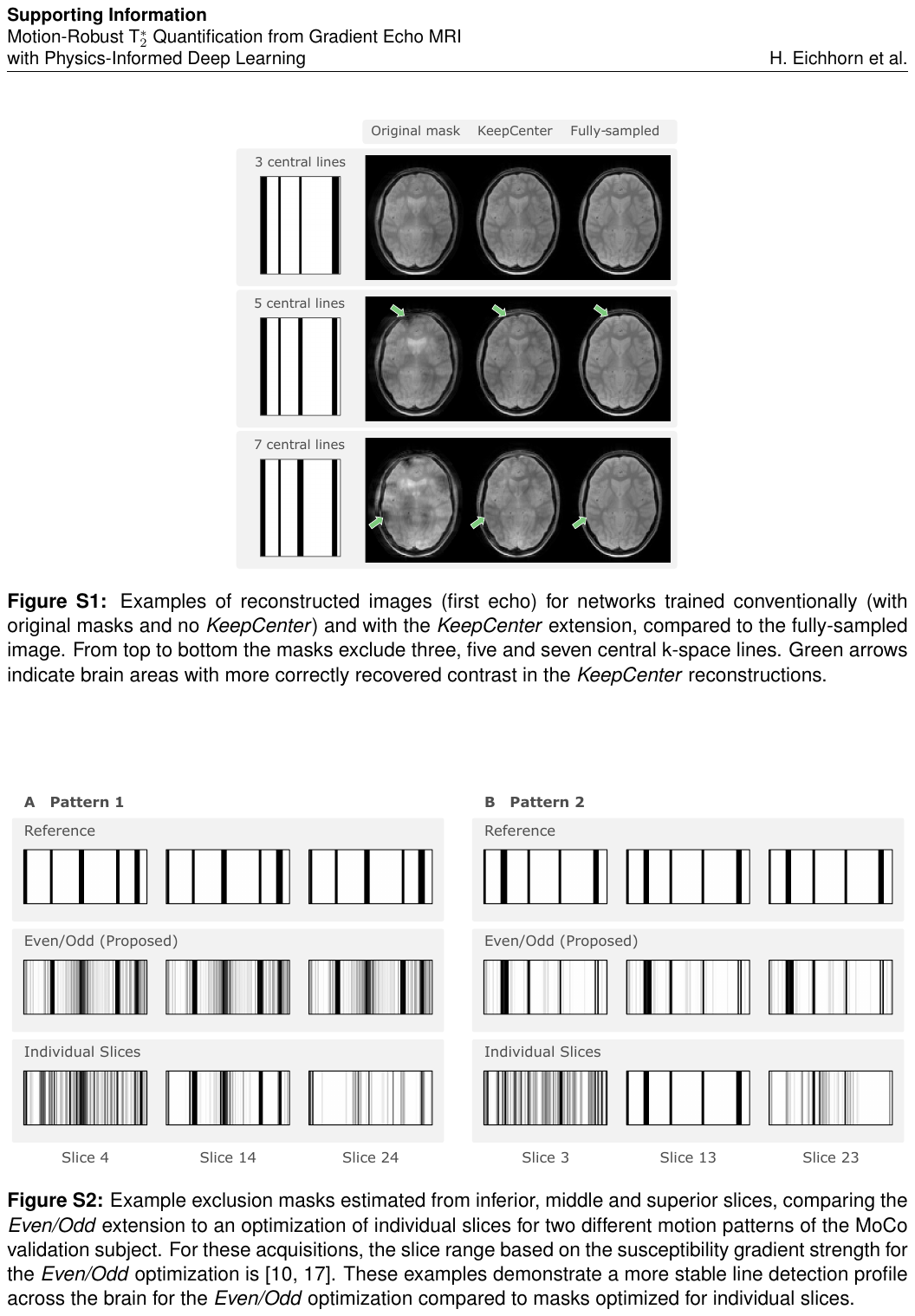}

\end{document}